\documentclass[twocolumn,showpacs,superscriptaddress,preprintnumbers,amsmath,amssymb,prc]{revtex4}

\usepackage{graphicx}
\usepackage{dcolumn}
\usepackage{bm}

\usepackage{color}

\begin{document}
\title{Thermal and transport properties in central heavy-ion reactions around a few hundred MeV/nucleon}
\author{X. G. Deng}
\affiliation{Shanghai Institute of Applied Physics, Chinese Academy
of Sciences, Shanghai 201800, China}
\affiliation{University of Chinese Academy of Sciences, Beijing 100049, China}
\author{Y. G. Ma\footnote{Corresponding author: ygma@sinap.ac.cn}}
\affiliation{Shanghai Institute of Applied Physics, Chinese Academy
of Sciences, Shanghai 201800, China}
\affiliation{ShanghaiTech University, Shanghai 200031, China}
\author{M. Veselsk\'{y}}
\affiliation{Institute of Physics, Slovak Academy of Sciences, D\'{u}bravsk\'{a} cesta 9,84511 Bratislava, Slovakia}

\date{\today}

\begin{abstract}

 Thermalization process of nuclear matter in central fireball region of heavy-ion collisions is investigated by employing an extension model of Boltzmann-Uehling-Uhlenbeck, namely the Van der Waals Boltzmann-Uehling-Uhlenbeck (VdWBUU) model.  Temperature ($T$) is extracted by the quantum Fermion fluctuation approach and other thermodynamic quantities, such as density ($\rho$), entropy density ($s$), shear viscosity ($\eta$), isospin diffusivity  ($D_{I}$) and heat conductivity ($\kappa$), are also deduced. %
 The liquid-like and gas-like phase signs are discussed through the behavior of shear viscosity during heavy-ion collisions process with the VdWBUU model.

\end{abstract}

\pacs{25.70.-z, 
      24.10.Lx, 
      21.30.Fe 
      }

\maketitle

\section{Introduction}
\label{introduction}

Heavy-ion collision  provides an unique tool for understanding properties of nuclear matter in different nuclear temperatures and densities, and determining the nuclear equation of state (EoS) \cite{BALI,TAMU}. Many observables are responsible for learning properties of the nuclear matter, such as thermodynamic variables and transport coefficients. A core observable is the nuclear temperature which has been extensively investigated by theories and experiments with different approaches~\cite{AK06} like double ratio of isotopic yield~\cite{JP95,JW05}, kinetic approaches~\cite{GDW82,BVJ83,GDW78,TO00,JG77,DGd02,RO04}, isospin thermometer~\cite{KHS93,KHS02}, double Fermi-sphere~\cite{DTK92}, classical fluctuation method~\cite{SW10} and quantum fluctuation method~\cite{HZ11,HZ13}. Nevertheless, there is no consensus on the best thermometer from the nuclear system~\cite{AK06}. One motivation to determine temperature is to investigate the liquid-gas phase transition in nuclear matter. In previous works~\cite{JP95,YGM97,JS97,BKS10,MaPRL,MaPRC,JBN}, many authors made efforts to study the liquid-gas phase transition in heavy-ion reactions. Classical liquid has a feature that shear viscosity decreases with the increasing of temperature  ~\cite{ENCA34,TWC66,JK78,LB89,MB92}. However, the situation of gas is the opposite~\cite{CB46,MB92}. For a micro-system, it is an excited subject to investigate shear viscosity of nuclear matter. In addition, the ratio of shear viscosity over entropy density ($\eta/s$) seems to have the bound of $\hbar/(4{\pi})$ which is proposed by Kovtun-Son-Starinets (KSS) in certain supersymmetric gauge theory~\cite{PKK05}. For years, attentions were paid to this value of quark-gluon matter produced in relativistic energies ~\cite{Nal,Gavin,Lacey,Maj,XuZhe,ND09,Nor,Luz,Bou,Song}. However, studies are very limited on shear viscosity of nuclear matter formed at lower energies \cite{NA09,Dang,LI11,ZHOU12,CLZ13,DQF14,ZHOU14,Pal,JX13}. In addition, in contrast with the viscosity coefficient, other transport coefficients like heat conductivity and isospin diffusion are still less mentioned for nuclear matter. Considering the above situations, our present work will focus on studies of transport coefficients in intermediate energy heavy-ion collisions.

The paper is organizied as follows: In Sec. \ref{VdW}, the simulation model and calculated formalism are introduced. In Sec. \ref{results}, thermal and transport results of central nuclear matter are extracted, and signals of liquid-like and gas-like phases are discussed by the temperature-dependent shear viscosity.
Finally, conclusion is made in Sec. \ref{summary}.

\section{The VdWBUU model and formalism}
\label{VdW}

\subsection{Van der Waals Boltzmann-Uehling-Uhlenbeck}

The Boltzmann-Uehling-Uhlenbeck (BUU) model is a popular tool for describing intermediate energy heavy-ion collisions~\cite{GFB84,HK85}. As a one-body mean-field theory based upon the Boltzmann equation~\cite{CYW78}, the BUU equation reads~\cite{WB86}:
\begin{eqnarray}
&&\frac{\partial f}{\partial t}+\upsilon\nabla_{r}f-\nabla_{r}U\nabla_{p}f   \notag\\
&&=\frac{4}{(2\pi)^{3}}\int d^{3}p_{2}d^{3}p_{3}
d\Omega\frac{d\sigma_{NN}}{d\Omega}\upsilon_{12}\times[f_{3}f_{4}(1-f)       \notag \\
&&\times(1\!-\!f_{2})\!-\!ff_{2}(1\!-\!f_{3})(1\!-\!f_{4})]\delta^{3}(p+p_{2}\!-\!p_{3}\!-\!p_{4}),
\label{BUU}
\end{eqnarray}%
where $f=f(r,p,t)$ is the phase-space distribution function, which can be solved using the method of Bertsch and Das Gupta~\cite{GFB88}; $\frac{d\sigma_{NN}}{d\Omega}$ is the in-medium nucleon-nucleon cross section; and $\upsilon_{12}$ is the relative velocity for colliding nucleons. $U$ is the mean-field potential including the isospin-dependent symmetry energy term,
\begin{eqnarray}
U(\rho,\tau_{z})=a(\frac{\rho}{\rho_{0}})+b(\frac{\rho}{\rho_{0}})^{\kappa}
+2a_{s}(\frac{\rho}{\rho_{0}})^{\gamma}\tau_{z}I,
\label{MFP}
\end{eqnarray}%
where $\rho_{0} = 0.168$ fm$^{-3}$ is the normal nuclear matter density; $I=(\rho_{n}-\rho_{p})/\rho$, with $\rho,\rho_{n} $and $\rho_{p}$ being densities of nucleons, neutrons and protons, respectively; $\tau_{z}=1$ for neutrons and $\tau_{z}=-1$ for protons; $a_{s}$ is coefficient of the symmetry energy term; $\gamma$ describes the density dependence; and $a, b,$ and $\kappa$ are parameters for the nuclear equation
of state (EoS). In this paper, we use two typical sets of mean-field parameters, the hard EoS with the compressibility $K$ of $380$ MeV ($a = -124$ MeV, $b = 70.5$ MeV, $\kappa = 2$) and the soft EoS with $K$ of $200$ MeV ($a = -356$ MeV, $b = 303$ MeV, $\kappa = 7/6$). In the model, the Coulomb interaction is also considered.

There are different versions of BUU model. The main differences are the extension of improvement potential and degree of freedom of isospin, which is explicitly taken into account~\cite{BAL97}. Here we use the version of Van der Waals Boltzmann-Uehling-Uhlenbeck model which was developed by Veselsk{\'y} and Ma ~\cite{MV13}. The pressure ($p$) changes in the thermodynamical equation of state, as a measure of nonideality of a neutron or a proton gas, can be defined as:
\begin{eqnarray}
 p = \rho^{2}\frac{\partial{\mathcal{U}}}{\partial{\rho}},
\label{pressureA}
\end{eqnarray}%
where $\mathcal{U}$ is the thermodynamic potential. When $\mathcal{U}$ is evaluated as sum of single-particle contributions of neutrons and protons, shown in Eq.(\ref{MFP}), Eq.(\ref{pressureB}) can be obtained:
\begin{eqnarray}
p = (\frac{f_{5/2}(z)}{f_{3/2}(z)})\rho{T}\!+\!a\rho^{2}
\!+\!b\kappa\rho^{1+\kappa}\!+\!2\gamma{a_{s}}\rho_{0}(\frac{\rho}{\rho_{0}})^{1+\gamma}\tau_{z}I,
\label{pressureB}
\end{eqnarray}%
where $z = exp(\mu/T)$ is the fugacity value of nucleons, with $\mu$ being chemical potential and $T$ is the temperature; and $\frac{f_{5/2}(z)}{f_{3/2}(z)}$ is a factor, a fraction of the Fermi integrals $f_{n}(z)=\frac{1}{\Gamma(n)}\int_{0}^{\infty}\frac{x^{n-1}}{z^{-1}e^{x}+1}dx$.
Based on the Van der Waals equation of state, using particle density $\rho$, one can obtain
\begin{eqnarray}
(p+a^{\prime}\rho^{2})(1-\rho{b^{\prime}})=(\frac{f_{5/2}(z)}{f_{3/2}(z)})\rho{T},
\label{VDW}
\end{eqnarray}%
where $a^{\prime}$ is related to attractive interaction among particles and $b^{\prime}$ denotes the proper volume of the constituent particle, which can be related geometrically to its cross section for interaction with other particles. Comparing Eq.(\ref{pressureB}) with Eq.(\ref{VDW}), one has:
\begin{eqnarray}
a^{\prime}&=&-a,                                                        \label{AAA}\\
b^{\prime}&=&\frac{b\kappa\rho^{\kappa}+2\gamma{a_{s}}(\frac{\rho}{\rho_{0}})^{1+\gamma}\tau_{z}I}
{(\frac{f_{5/2}(z)}{f_{3/2}(z)})\rho{T}+b\kappa\rho^{1+\kappa}
+2\gamma{a_{s}}\rho_{0}(\frac{\rho}{\rho_{0}})^{1+\gamma}\tau_{z}I}.    \label{BBB}
\end{eqnarray}%
In a system constituting with nucleons, the proper volume can be used to estimate its cross section within the nucleonic medium
\begin{eqnarray}
\sigma=(\frac{9\pi}{16})^{1/3}b^{\prime{2/3}},
\label{CROSS}
\end{eqnarray}%
which can be implemented into the collision term of the Boltzmann-Uehling-Uhlenbeck equation. In this way, the Boltzmann-Uehling-Uhlenbeck can be simulated with both isospin-dependent mean-field and nucleon-nucleon cross sections, which are correlated to each other.  This is a so-called  VdWBUU equation. More details can be found in Ref.~\cite{MV13}.

\begin{figure*}[htb]
\centerline{
\includegraphics[scale=0.97]{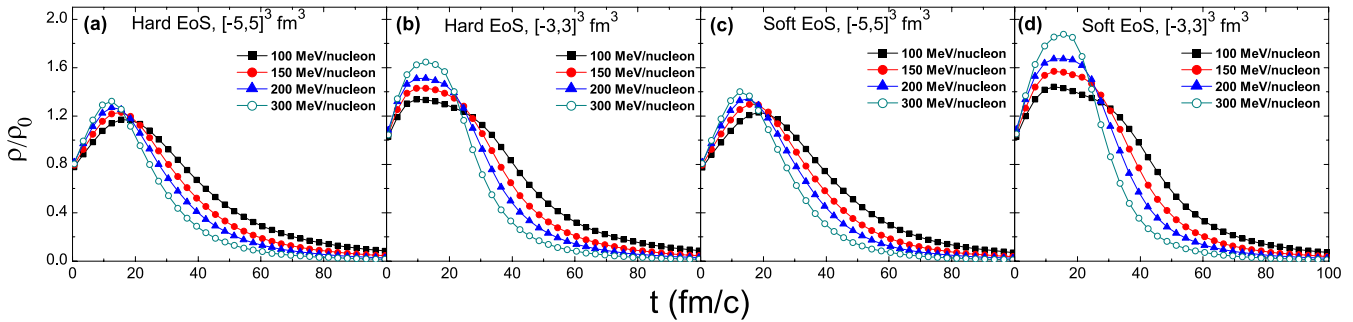}}
 \caption{(Color online) Time evolution of average density at different incident energies for: (a) the hard EoS in the region of $X$[-5,5], $Y$[-5,5], $Z$[-5,5]; (b) the hard EoS in the region of $X$[-3,3], $Y$[-3,3], $Z$[-3,3]; (c) the soft EoS in the region of $X$[-5,5], $Y$[-5,5], $Z$[-5,5]; (d) the soft EoS in the region of $X$[-3,3], $Y$[-3,3], $Z$[-3,3].}
\label{F1}
\end{figure*}

\subsection{Evaluated formalism}

In this paper, thermodynamic and transport properties of a nuclear system are investigated in the framework of the VdWBUU approach. Thermodynamic quantities are extracted with different formalisms.

The concept of nuclear temperature was proposed by Bethe~\cite{HAB37} and Weisskopf~\cite{VFW37}. Different thermometer methods to extract nuclear temperature are given in Ref.~\cite{AK06}. In Refs.~\cite{DTK92,CLZ13}, the hot Thomas-Fermi formalism was used to calculate the temperature. In this paper, a method based on quantum fluctuation of Fermions is applied to calculate the temperature, proposed in Ref.~\cite{HZ11}. The quadrupole $Q_{xy}=p_{x}^{2}-p_{y}^{2}$ is defined and the variance $\langle\sigma_{xy}^{2}\rangle$ is given in Ref.~\cite{SW10}:
\begin{eqnarray}
\langle\sigma_{xy}^{2}\rangle=\int{d^3p({p_{x}^{2}}-{p_{y}^{2}})^2}n(p),
\label{variance}
\end{eqnarray}%
where $n(p)$ is the momentum distribution of particles. In heavy-ion collisions, protons, neutrons and tritium follow the Fermi statistics, thus the Fermi-Dirac distribution can be used in Eq.(\ref{variance})~\cite{SW10}. Using the Fermi-Dirac distribution $n(p)$, we have
\begin{eqnarray}
\langle\sigma_{xy}^{2}\rangle&=&\frac{\int{d^3p({p_{x}^{2}}-{p_{y}^{2}})^2}n(p)}{\int{d^3p}n(p)}   \notag \\
&=&(2mT)^{2}\frac{4}{15}\frac{\int_{0}^{\infty}y^{\frac{5}{2}}\frac{1}{z^{-1}e^{y}+1}dy}
{\int_{0}^{\infty}y^{\frac{1}{2}}\frac{1}{z^{-1}e^{y}+1}dy}                                          \notag \\
&=&(2mT)^{2}F_{QC}(z),
\label{varianceA}
\end{eqnarray}%
where $z = exp(\mu/T)$, and $F_{QC}(z)$ is the quantum correction factor. Chemical potential ($\mu$) can be extracted by the following equation, namely ~\cite{DQF14}:
\begin{eqnarray}
\rho_{\tau}&=&\frac{g}{(2\pi\hbar)^3}{\int}n_{\tau}(p)d^3p \notag \\
&=&\frac{1}{2\pi^{2}}(\frac{2m}{\hbar^{2}})^{\frac{3}{2}}\int_{0}^{\infty}\frac{\sqrt{y}}{z_{\tau}^{-1}e^{y}+1}dy.
\label{chemical}
\end{eqnarray}%
Here $g(=2)$ is the spin degeneracy of nucleon and $\tau$ denotes $n$ for neutron or $p$ for proton. With the variance $\langle\sigma_{xy}^{2}\rangle$, chemical potential and density calculated by data from spatial distribution of test particles, temperature of the nuclear system can be extracted.

Entropy density can be obtained by given density and temperature~\cite{HZ12}:
\begin{eqnarray}
s=\frac{U-A}{T}\frac{1}{V}=[\frac{5}{2}\frac{f_{5/2}(z)}{f_{3/2}(z)}-\ln{z}]\rho,
\label{entropy}
\end{eqnarray}%
where $U$ is the internal energy and $A$ is Helmholtz free energy~\cite{LL80}.

In order to learn more about the properties of  hot nuclei during heavy-ion collisions or nuclear EoS, the transport coefficients are evaluated, including shear viscosity ($\eta$), isospin diffusivity ($D_{I}$) and heat conductivity ($\kappa$). In hydrodynamics, the internal friction occurs when there exists relative motions  in a fluid (liquid or gas), and this is called viscosity. The shear viscosity depends on many factors of fluid species, velocity gradient, temperature and density. Diffusivity is a measure of the rate at which particles or heat or fluids can spread. Here, isospin diffusivity represents the ability of isospin diffusion in nuclear matter. The heat conductivity represents the ability of a material to conduct heat. These three transport coefficients are important for the nucleonic transport process of nuclear matter and were discussed in Ref.~\cite{LS03} for a two-component nuclear Fermi system. By solving the Boltzmann-equation set such as used in the reaction simulations~\cite{GFB88},
the numerical results for the coefficients have been obtained,
\begin{eqnarray}
\eta(\rho,T,I)&=&(1+0.10I^{2})[\frac{856}{T^{1.10}}(\frac{\rho}{\rho_0})^{1.81}
  \notag \\ &-&
\frac{240.9}{T^{0.95}}(\frac{\rho}{\rho_0})^{2.12}
+2.154T^{0.75}],                                                                   \label{coefficientA}\\
D_{I}(\rho,T,I)&=&(1-0.19I^{2})[\frac{11.34}{T^{2.38}}(\frac{\rho}{\rho_0})^{1.54}
 +\frac{1.746}{T}(\frac{\rho}{\rho_0})^{0.56}
  \notag \\
&+&0.00585T^{0.913}(\frac{\rho}{\rho_0})],                                             \label{coefficientB}\\
\kappa(\rho,T,I)&=&(1+0.10I^{2})[\frac{0.235}{T^{0.755}}(\frac{\rho}{\rho_0})^{0.951}
 \notag \\
 &-&0.0582(\frac{\rho}{\rho_0})^{0.0816}                                                   \notag \\
&+&0.0238T^{0.5627}(\frac{\rho}{\rho_0})^{0.0171}].                                    \label{coefficientC}
\end{eqnarray}%
where $T$ is temperature in MeV, $\eta$ is shear viscosity in MeV/(fm$^{2}$c), $D_{I}$ is isospin diffusivity in fm$\cdot$c, heat conductivity $\kappa$ is in c/fm$^{2}$; and $I=\frac{\rho_{n}-\rho_{p}}{\rho}$ is isospin asymmetry. More details can be found  in Ref.~\cite{LS03}. The numerical results for these coefficients are calculated by using the experimentally measured nucleon-nucleon cross section as $\sigma_{free}=40 mb$.  More accurately, we have to modify these transport coefficients by the in-medium N-N cross sections ($\sigma$) as we adopted in Ref.~\cite{CLZ13}
 which can be extracted with  Eqs.(\ref{BBB}) and (\ref{CROSS}).
\begin{eqnarray}
\eta(\rho,T,I,\sigma)&=&\frac{\eta(\rho,T,I)}{\sigma/\sigma_{free}},
\label{coefficientAC}\\
D_{I}(\rho,T,I,\sigma)&=&\frac{D_{I}(\rho,T,I)}{\sigma/\sigma_{free}},
\label{coefficientBC}\\
\kappa(\rho,T,I,\sigma)&=&\frac{\kappa(\rho,T,I)}{\sigma/\sigma_{free}}.
\label{coefficientCC}
\end{eqnarray}%

\section{Results and discussion}
\label{results}

Central collisions (b=0 fm) of $^{197}$Au+$^{197}$Au are simulated at beam energies of 100-300 MeV/nucleon, employing the VdWBUU model with the hard EoS and soft EoS. The central region is defined as a $[-5,5]^3$ fm$^3$ or $[-3,3]^3$ fm$^3$ box and its center located in the c.m. We denote that $t=0$ fm/c is at the point where two nuclei touch initially.

\subsection{Properties of thermodynamic quantities}
\label{resultsA}

\begin{figure*}[htb]
\setlength{\abovecaptionskip}{0pt}
\setlength{\belowcaptionskip}{8pt}\centerline{\includegraphics[scale=0.97]{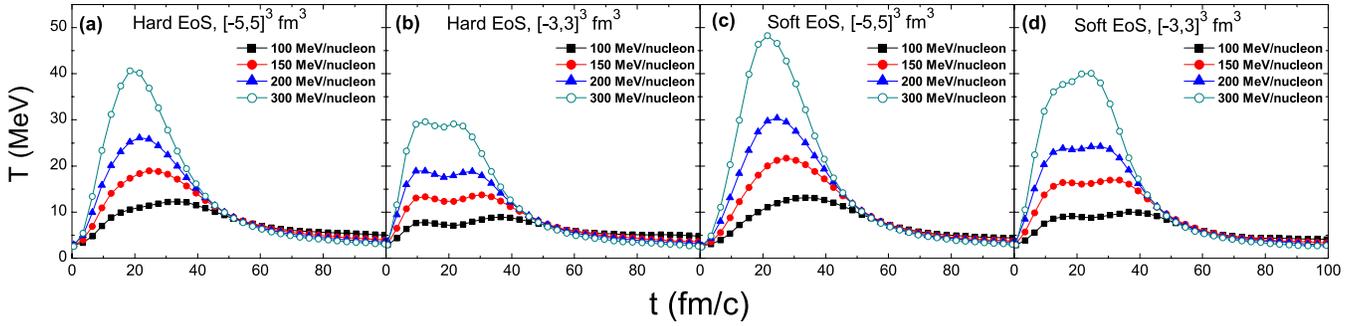}} \caption{(Color
online) Time evolution of average temperature at different incident energies. The conditions are the same as   in Fig.~\ref{F1}.}
\label{F3}
\end{figure*}

\begin{figure}[htb]
\setlength{\abovecaptionskip}{0pt}
\setlength{\belowcaptionskip}{8pt}\centerline{\includegraphics[scale=0.58]{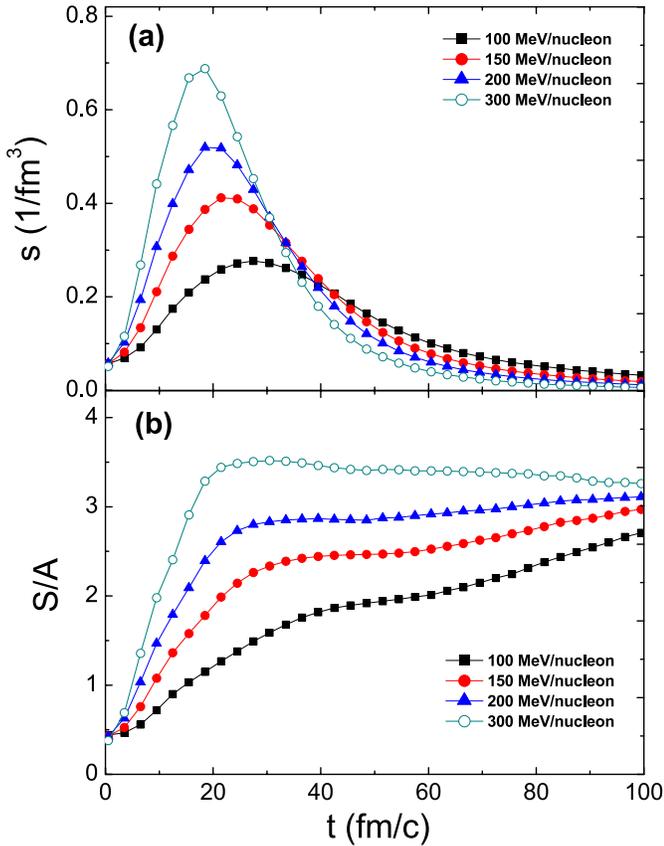}}
\caption{(Color online) Time evolution of average entropy density (a) and entropy per nucleon (b) at different incident energies for the soft EoS in the region of $X$[-5,5], $Y$[-5,5] and $Z$[-5,5]. }
\label{F4}
\end{figure}

\begin{figure}[htb]
\setlength{\abovecaptionskip}{0pt}
\setlength{\belowcaptionskip}{8pt}\centerline{\includegraphics[scale=0.64]{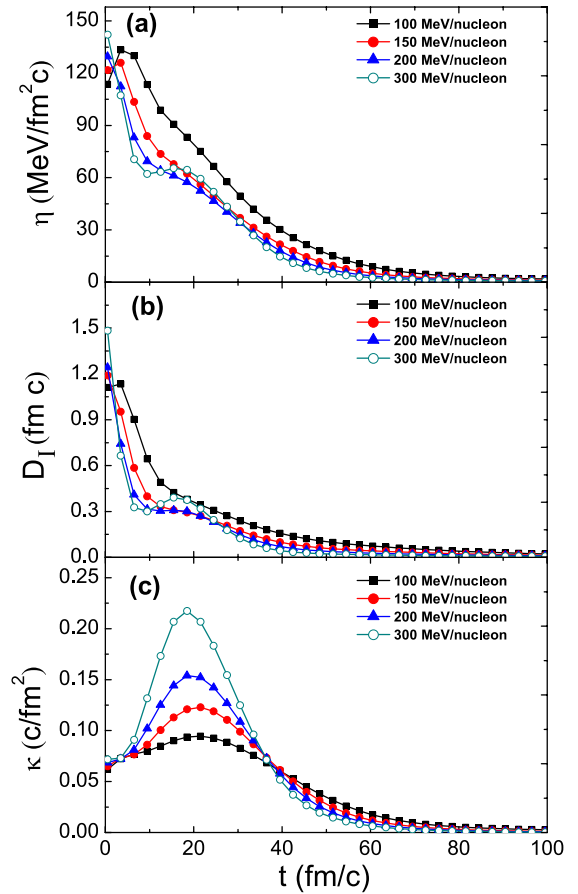}}
 \caption{(Color online) Time evolution of shear viscosity (a), isospin diffusivity (b) and heat conductivity (c) at different incident energies for the soft EoS in the region of $X$[-5,5], $Y$[-5,5] and $Z$[-5,5].}
\label{F5}
\end{figure}

\begin{figure}[htb]
\setlength{\abovecaptionskip}{0pt}
\setlength{\belowcaptionskip}{8pt}
\centerline{\includegraphics[scale=0.45]{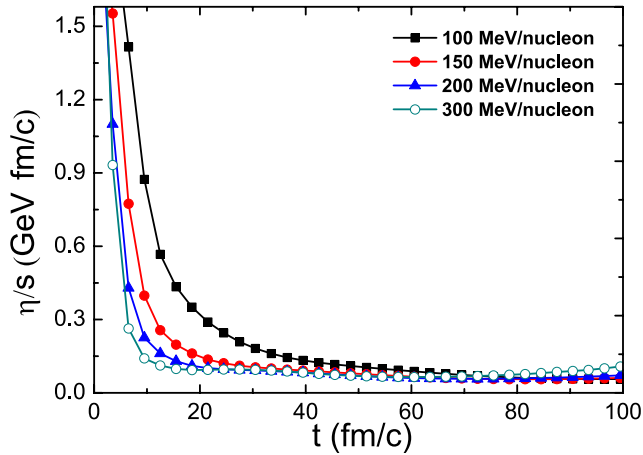}}
\caption{(Color online) Time evolution of the ratio of shear viscosity over entropy density at different incident energies for the soft EoS in the region of $X$[-5,5], $Y$[-5,5] and $Z$[-5,5]. }
\label{SVstime}
\end{figure}

\begin{figure}[htb]
\setlength{\abovecaptionskip}{0pt}
\setlength{\belowcaptionskip}{8pt}
\centerline{\includegraphics[scale=0.55]{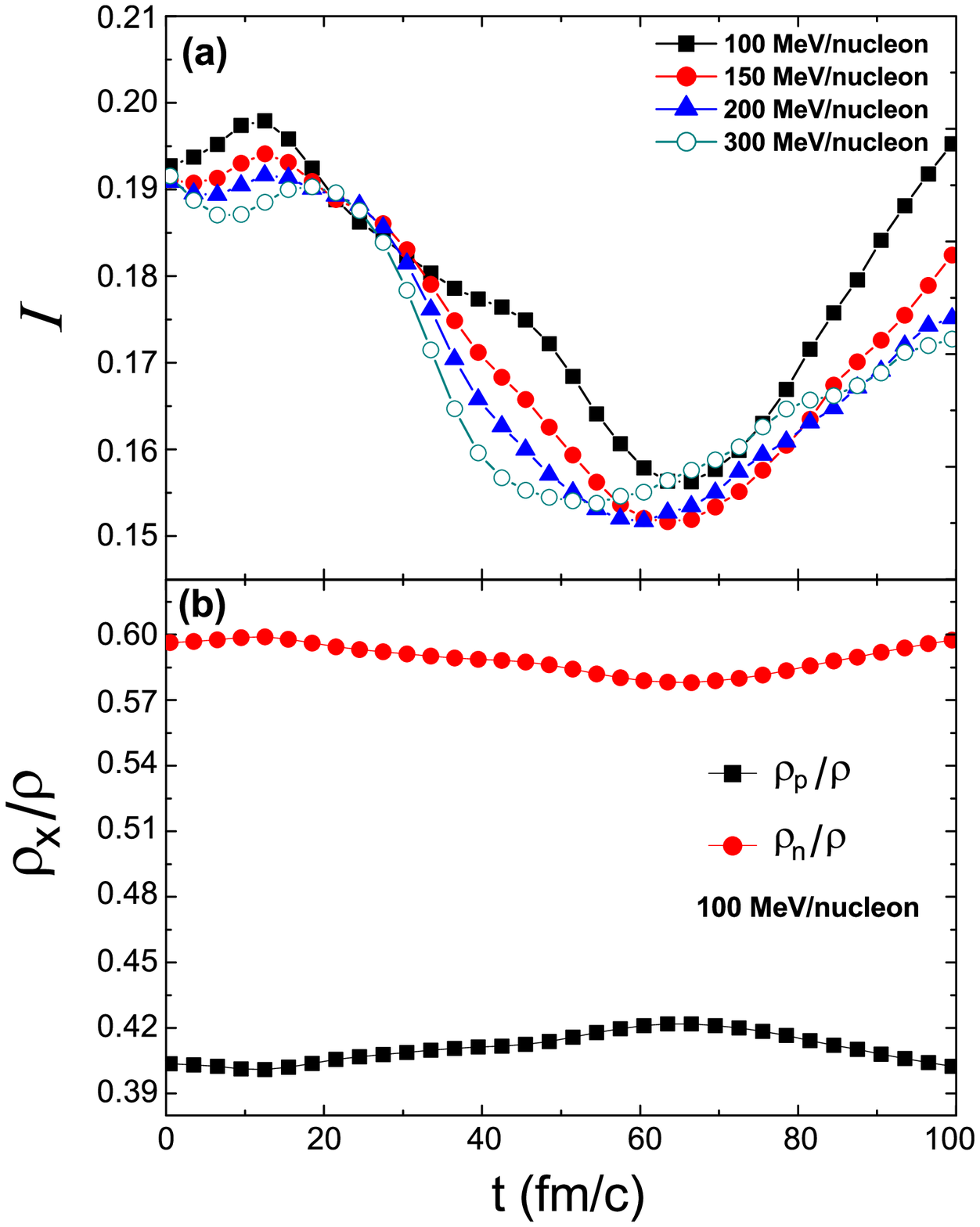}}
\caption{(Color online) Time evolution of isospin asymmetry at different incident energies (a) and various reduced densities ($\rho_n$, $\rho_p$ and $\rho$ represent the density of neutron, proton and the total, respectively; x is n or p.) at 100 MeV/nucleon (b) for the soft EoS in the region of $X$[-5,5], $Y$[-5,5] and $Z$[-5,5].}
\label{Irho}
\end{figure}

Fig.~\ref{F1} shows the time evolution of average density in the central region $[-5,5]^3$ fm$^3$ or $[-3,3]^3$ fm$^3$  at incident energies of $100-300$ MeV/nucleon with the hard or soft EoS. One sees that the average density is $1.2-1.6$ times of normal nuclear matter density, with larger maxima at earlier time in higher beam energies. At the expansion stage, higher beam energies lead rapidly to lower densities. At the same incident energy, comparing Fig.~\ref{F1}(a) with Fig.~\ref{F1}(c), densities of the hard EoS and soft EoS are similar; while comparing Fig.~\ref{F1}(a) with Fig.~\ref{F1}(b), or Fig.~\ref{F1}(c) with Fig.~\ref{F1}(d), the density is higher in the smaller region at the maximum point. The reason is that the density of the zone far from the center is lower than the close one's.

The time evolution of average temperature is shown in Fig.~\ref{F3}. In the initial stage the average temperature is nearly zero because of two nuclei being cold with Fermi momenta at the beginning of collision. As two nuclei come close to each other, with increasing nucleon-nucleon collisions, the temperature increases and reaches the maximum of $13-48$ MeV, and it is higher at higher incident energies. At the expansion stage, the temperature decreases and falls  faster at higher incident energies. By comparing Fig.~\ref{F3}(a) with Fig.~\ref{F3}(c), the maximum temperature with the hard EoS is lower than that of the soft EoS. This indicates that the collision with the soft EoS is compressed more easily than with the hard one and more beam motions are transformed into thermal motions. Along the time scale of the collision, the maximum points of temperature at different beam energies in Fig.~\ref{F3}(a) or Fig.~\ref{F3}(c) delay in comparison with those of density in Fig.~\ref{F1}(a) or Fig.~\ref{F1}(c)), indicating that the nucleon-nucleon collisions are still frequent when the maximum compression reached. The features of Fig.~\ref{F3}(c) are similar to those of Zhou {\it et al.}~\cite{CLZ13} who used the Thomas-Fermi method to obtain temperature with the isospin dependent quantum molecular dynamics (IQMD) model for  a central spherical region with radius of $5$ fm. Of course, the quantitative value of temperature relies on the model and thermometer. Meanwhile, in Figs.~\ref{F3}(b) and ~\ref{F3}(d) in the $[-3,3]^3$ fm$^3$ region, the average temperature keeps saturated near the peak point. The size effect is obvious. A smaller central region has fewer nucleons, thus more interaction information may not be included. So if one wants to extract the information from central region, the suitable size should be considered. To simplify the structure of the present paper, we just present the following calculation results with the soft EoS. Of course, the case with the hard EoS keeps the similar qualitative behaviors.

Time evolutions of average entropy density and entropy per nucleon are shown in Fig.~\ref{F4}. Since  the system is an open system with non-fixed nucleon numbers, the information of  entropy per nucleon shall be interesting. During the compression process between two nuclei, nucleon number in the region and state numbers of the system will increase naturally, hence the increase of average entropy density. Conversely, it decreases at the expansion stage. The time evolution of average entropy density is similar to the case of  average temperature, i.e. higher beam energy leads to larger entropy density at the maximum value of $0.1-0.7$ fm$^{-3}$. This shows that the higher beam energy with the larger internal energy in the central region, the more state numbers and entropy will be. However, at different incident energies, the average entropy densities reach to the maximum earlier than those of average temperatures (to be discussed later). At the expansion stage, average entropy density decreases. And entropy per nucleon shows more or less saturation at higher beam energy, but it shows a slight increase with time at lower one, indicating the potential of the open system converts slowly to internal energy after maximum compression for the lower incident energy. In other words, it means the central fireball tends to equilibration faster at higher incident energy.

\begin{figure}[htb]
\setlength{\abovecaptionskip}{0pt}
\setlength{\belowcaptionskip}{8pt}
\centerline{\includegraphics[scale=0.45]{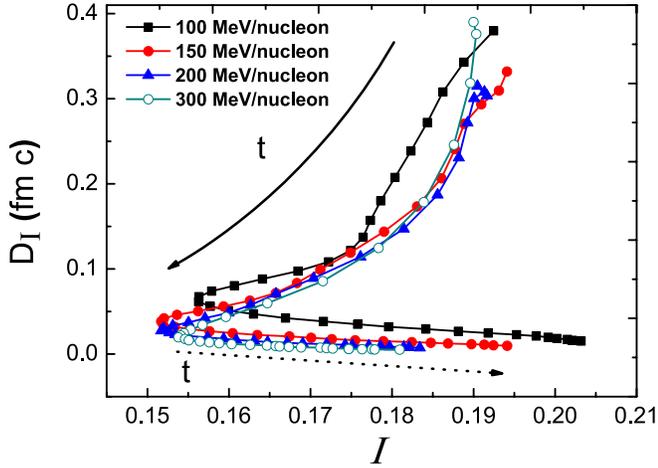}} \caption{(Color
online) Isospin diffusivity as a function of isospin asymmetry at $t=15-100$ fm/c at different incident energies for the soft EoS in the region of $X$[-5,5], $Y$[-5,5] and $Z$[-5,5]. The solid and dotted arrows indicate former and later process of expansion, respectively. }
\label{DItimesw}
\end{figure}

\begin{figure}[htb]
\setlength{\abovecaptionskip}{0pt}
\setlength{\belowcaptionskip}{8pt}\centerline{\includegraphics[scale=0.45]{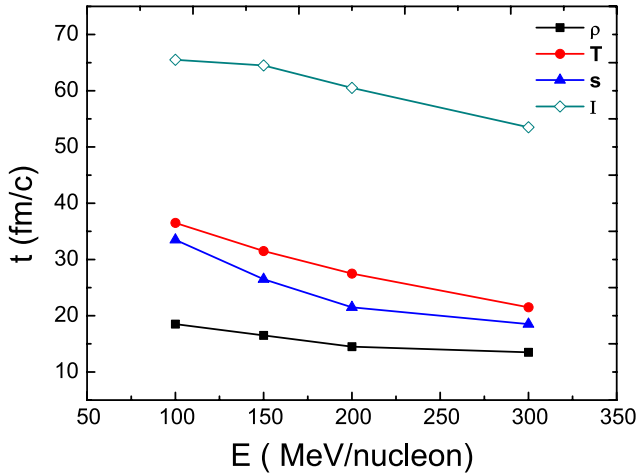}} \caption{(Color
online) The time points of maximum average density (solid squares),  maximum average entropy density (open triangle), maximum average temperature (solid circle), and minimum isospin asymmetry (open diamond) as a function of beam energy for the soft EoS in the region of $X$[-5,5], $Y$[-5,5] and $Z$[-5,5]. }
\label{Mtime}
\end{figure}

\begin{figure}[h]
\setlength{\abovecaptionskip}{0pt}
\setlength{\belowcaptionskip}{8pt}\centerline{\includegraphics[scale=0.45]{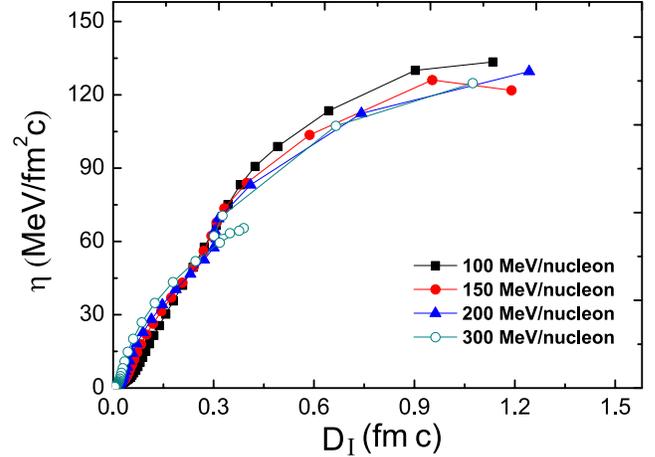}}
\caption{(Color online) Shear viscosity as a function of isospin diffusivity at different incident energies for the soft EoS in the region of $X$[-5,5], $Y$[-5,5], $Z$[-5,5]. }
\label{F6}
\end{figure}

\begin{figure}[htb]
\setlength{\abovecaptionskip}{0pt}
\setlength{\belowcaptionskip}{8pt}\centerline{\includegraphics[scale=0.45]{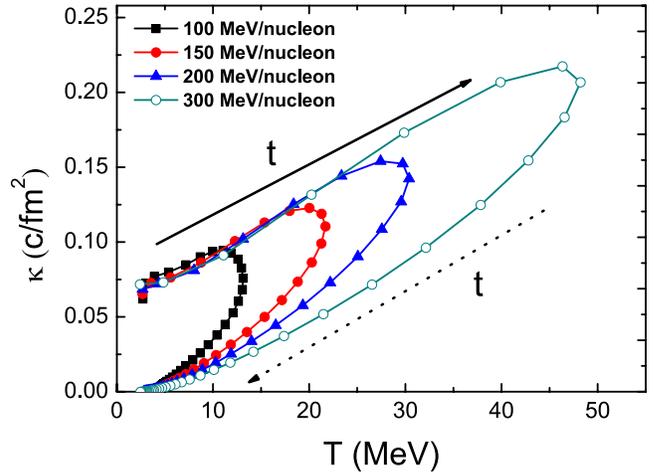}} \caption{(Color
online) Heat conductivity as a function of temperature at different incident energies for the soft EoS in the region of $X$[-5,5], $Y$[-5,5] and $Z$[-5,5]. The solid and dotted arrows indicate compression process and expansion process, respectively.}
\label{KT}
\end{figure}

Shear viscosity, isospin diffusivity and heat conductivity are shown in Fig.~\ref{F5}. Shear viscosity, isospin diffusivity and heat conductivity decrease drastically in the initial compression stage ($0-20$ fm/c). Shear viscosity and isospin diffusivity tend to an asymptotic value after $20$ fm/c, respectively, while heat conductivity shows a peak around $20$ fm/c, indicating the compressed nuclear matter has stronger heat conduct capability at the stage of higher temperature and higher density.

In Fig.~\ref{SVstime}, the ratio of shear viscosity over entropy density  also decreases drastically in the initial compression stage ($0-20$ fm/c) and tends to an asymptotic value after $20$ fm/c. It is noted that the smaller value of the ratio of shear viscosity over entropy density is found at higher incident energy during early time evolution. Overall, $\eta/s$ values approximate to asymptotic  values above  the KSS bound of $\sim \frac{\hbar}{4\pi}$ at the expansion stage.

In Fig.~\ref{Irho}(a), isospin asymmetry ($I$) changes little before $20$ fm/c but drops to a minimum at  $\sim 60$ fm/c and then increases again. At lower energies, the valley of $I$ is a relative shallow.
This valley is associated with the density evolutions of neutrons and protons. For instance, the evolutions of reduced density of neutrons and  protons over  the total density at incident energy of $100$ MeV/nucleon are shown in Fig.~\ref{Irho}(b). Obviously, the reduced  density of neutrons generally decreases before $\sim 60$ fm/c and vice versa for the reduced  density of protons.  Therefore, the isospin, defined as ($\rho_{n}-\rho_{p}$)/$\rho$, shows a dip around 60 fm/c in Fig.~\ref{Irho}(a). This dip point induces an inflection point of isospin diffusivity versus  isospin asymmetry as shown in Fig.~\ref{DItimesw}.

In above figures, peaks or valleys of some quantities observed during their time evolution are more or less related to the transformation from the compression stage to the expansion stage in heavy-ion collisions. For comparing  the corresponding times of which the peaks or valleys for various observables, such as  density, temperature, entropy density, and isospin asymmetry reaching to the maximum or minimum, we can compare different time order of those quantities. Fig.~\ref{Mtime} shows the corresponding times at the  peak or valley of these quantities as a function of beam energy.
One can roughly see that every above quantity
decreases with increasing incident energy. Note that, the time point when the  temperature reaches later to its maximum at different incident energies than those of maximum density it reaches. This indicates that even though when the maximum compressed state is reached, the frequent nucleon-nucleon collisions are still going on  and therefore the system is still heating. The time points when the  entropy density  reaches its maximum are in between those of maximum density and maximum  temperature. It indicates that when the system reaches the maximum compression stage in the central region, internal energy increases with the $NN$ collision leading to the increasing numbers of  states and therefor the entropy density is still rising up at that time. In contrast, isospin asymmetry is much later for reaching its minimum in comparison with other observables since it is mostly due to isospin transport process not directly relates to compression-expansion process.

 As shown in Fig.~\ref{F6}, the isospin diffusivity shows a monotonic dependence on shear viscosity, indicating that the viscous nuclear matter is favor of the isospin diffusion. In Fig.~\ref{KT}, the heat conductivity increases as the temperature at compression stage which is similar to a classical behavior~\cite{LS03}. Last, heat conductivity drops at expansion stage. During compression stage, more frequent $NN$ collision leads to hotter and larger heat conductivity of the system. At the expansion stage, however, the system cools down with smaller heat conductivity.

\begin{figure}[htb]
\setlength{\abovecaptionskip}{0pt}
\setlength{\belowcaptionskip}{8pt}\centerline{\includegraphics[scale=0.45]{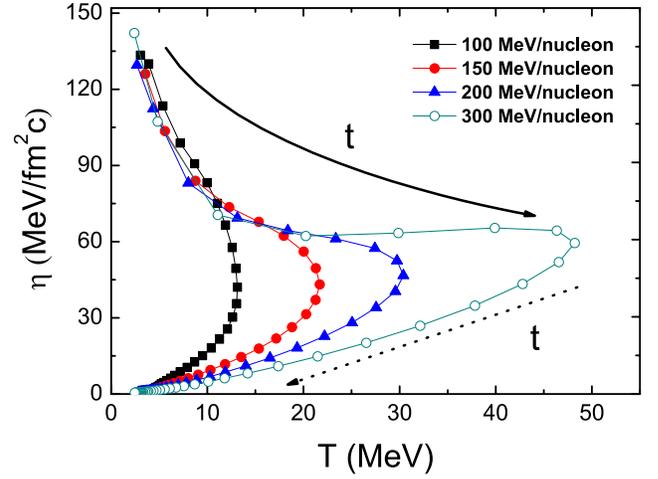}} \caption{(Color
online) Shear viscosity as a function of temperature at different incident energies for the soft EoS in the region of $X$[-5,5], $Y$[-5,5] and $Z$[-5,5]. The solid and dotted arrows indicate compression process and expansion process, respectively.}
\label{F7}
\end{figure}

\subsection{The signature of liquid-like and gas-like phases}
\label{resultsB}

The relationship between shear viscosity and temperature in liquid and gas of common substance differs from each other. In the present study on nuclear matter system, we find the similar behaviors of liquid-like and gas-like phases from the quantities we extracted. It should be mentioned that, in Fig.~\ref{F7}, the branch above the inflection point at different beam energies indicates the compression process, and the bottom branch is in the expansion process. One can find that in the compression process, shear viscosity decreases with increasing temperature; while in the expansion process the behavior is inverse. We also notice that in classical liquid, as the temperature goes higher,
the decrease of shear viscosity~\cite{TWC66} also appears. On the contrary, the shear viscosity of gas~\cite{CB46} and meson gas~\cite{AM04} drops with the decrease temperature as collision probability becomes smaller. More information can be obtained in Refs.~\cite{ENCA34,LB89,LD71,JK78,SY07}. In this context, it can be analogous that nuclear matter is in the liquid-like phase state during compression process and while it is gas-like phase during the later expansion process. Around the inflection point the system is a kind of mixed-like phase.

\begin{figure}[htb]
\centerline{\includegraphics[scale=0.45]{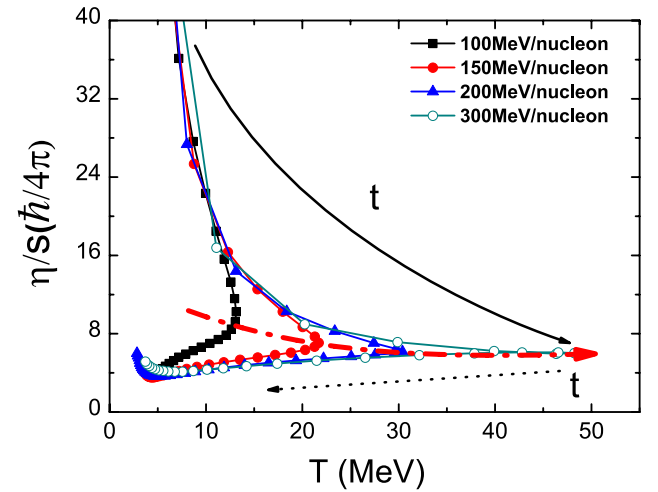}}
\caption{(Color
online) Ratio of the shear viscosity over entropy density, in units of $\frac{\hbar}{4\pi}$, as a function of temperature at different incident energies for the soft EoS in the region of $X$[-5,5],$Y$[-5,5 and $Z$[-5,5]. Red dashed line shows the trend of value of $\eta/s$ at the turning point with the temperature. }
\label{F9}
\end{figure}

\subsection{$\eta/s$ versus temperature or beam energy}
\label{resultsC}
As shown in Fig.~\ref{SVstime}, $\eta/s$ reaches to an asymptotic value which is above the KSS bound in the later stage when the system is fully expanded. Fig.~\ref{F9} shows the ratio of shear viscosity over entropy density as a function of temperature. It is noted that when the system reaches to the maximum temperature, there exists a turning point of $\eta/s$. This behavior is qualitatively consistent with the feature of $\eta$ as a function of temperature (Fig.~\ref{F7}).
Furthermore, if we observe the behavior of $\eta/s$ values at the point of maximum temperature (red dashed line in the figure), it displays that these ratios show fast drop at higher beam energy and reaches a plateau about $6$ times of the KSS bound as other model shows \cite{ZHOU12}.

\section{Conclusions}
\label{summary}
In summary, thermal and transport quantities for the nuclear matter formed in central Au + Au collisions at a few hundred MeV/nucleon are obtained from the VdWBUU model. The properties of central region of nuclear reaction are discussed with some quantities for different fireball sizes and nuclear equations of state. Time evolutions of  density, temperature, entropy density,  isospin diffusivity, shear viscosity, and heat conductivity etc. are presented, which give information on nuclear matter of collision system. The peak or valley behavior in the time evolution of density, temperature, entropy,
and isospin are more or less related to the compression and expansion process in heavy-ion collisions, with different time delays for the above different observables  after the most compressed state. Time order of the quantities have been displayed. The sign of liquid-like and gas-like phases is given by relations of the shear viscosity as a function of temperature.
The values of $\eta/s$ at the maximum temperature for different beam energies demonstrates a decrease behavior and tends to an asymptotic values of $\sim$ 6$\frac{\hbar}{4\pi}$.

\begin{acknowledgments}

This work was supported in part by the
National Natural Science Foundation of China under contract Nos. 11421505,
and 11220101005, and the Major State Basic Research
Development Program in China under Contract No. 2014CB845401. M. V. thanks the support by Chinese Academy of CAS President's International Fellowship Initiative No. 2011T2J13 and  the Slovak Research and Development Agency under contract APVV-15-0225.

\end{acknowledgments}

\end{document}